\documentclass{jfm}
\usepackage{graphicx}
\usepackage{epstopdf, epsfig}
\usepackage{amsmath}
\usepackage{subcaption}

\usepackage{color}
\usepackage{tikz}
\newcommand{\red}{\textcolor{black}}

\usepackage{layouts}
\usepackage[colorlinks = true,allcolors = {blue}]{hyperref}

\newcommand{\redline}{\raisebox{2pt}{\tikz{\draw[-,red,solid,line width = 1.5pt](0,0) -- (5mm,0);}}}
\newcommand{\blueline}{\raisebox{2pt}{\tikz{\draw[-,blue,solid,line width = 1.5pt](0,0) -- (5mm,0);}}}

\shorttitle{Periodically driven TC turbulence}
\shortauthor{R. A. Verschoof, A. K. te Nijenhuis, S. G. Huisman, C. Sun, D. Lohse}

\title{Periodically driven Taylor-Couette turbulence}

\author{Ruben A. Verschoof\aff{1}\thanks{These authors contributed equally to this work},
  Arne K. te Nijenhuis\aff{1}$\dagger$,
  Sander G. Huisman\aff{1},
 Chao Sun\aff{2,1}
  \corresp{\email{chaosun@tsinghua.edu.cn}},
\and Detlef Lohse\aff{1,3}
 \corresp{\email{d.lohse@utwente.nl}} 
 }
\affiliation{
\aff{1}Physics of Fluids, Max Planck Institute for Complex Fluid Dynamics, MESA+ Institute and J. M. Burgers Center for Fluid Dynamics, University of Twente, P.O. Box 217, 7500 AE Enschede, The Netherlands
\aff{2}Center for Combustion Energy and Department of Thermal Engineering, Tsinghua University, 100084 Beijing, China
\aff{3}Max Planck Institute for Dynamics and Self-Organization, Am Fassberg 17, 37077 G\"{o}ttingen, Germany}

\begin{document}

\maketitle

\begin{abstract}
We study periodically driven Taylor-Couette turbulence, i.e.\ the flow confined between two concentric, independently rotating cylinders. Here, the inner cylinder is driven sinusoidally while  the outer cylinder is kept at rest (time-averaged Reynolds number is $Re_i = 5 \times 10^5$). Using particle image velocimetry (PIV), we measure the velocity over a wide range of modulation periods, corresponding to a change in Womersley number in the range $15 \leq W\!o \leq 114$.  To understand how the flow responds to a given modulation, we calculate the phase delay and amplitude response of the azimuthal velocity.

In agreement with earlier theoretical and numerical work, we find that for large modulation periods the system follows the given modulation of the driving, i.e.\ the system behaves quasi-stationary. For smaller modulation periods, the flow cannot follow the modulation, and the flow velocity responds with a phase delay and a smaller amplitude response to the given modulation. If we compare our results with numerical and theoretical results for the laminar case, we find that the scalings of the phase delay and the amplitude response are similar. However, the local response in the bulk of the flow is independent of the distance to the modulated boundary. Apparently, the turbulent mixing is strong enough to prevent the flow from having radius-dependent responses to the given modulation. 
\end{abstract}


\section{Introduction}
Periodically driven turbulent flows are omnipresent. Well-known examples include blood flow driven by the beating heart, the flow in internal combustion engines, the earth's atmosphere which is periodically heated by the sun, and tidal currents caused by periodic changes in the gravitational attraction of both the moon and sun.  

One line of research assumes homogeneous isotropic turbulence. These studies focussed on the {\it global response} of the system, i.e.\ the response amplitude and the phase shift of the quantities such as a global Reynolds number \citep{loh00}, or the total energy in the system \citep{hey03a}. Most numerical studies in addition only used simplified models, such as the GOY shell model or the reduced wave vector set approximation (REWA) \citep{hoo01,hey03b,Hamlington2009}.  Only a limited number of DNS studies have been performed in this field, because of the computational costs needed to achieve both fully developed turbulence and sufficient statistical convergence with temporal dependence \citep{yu06,kuc06,Kuczaj2008}. Also studies on periodically driven wind tunnels were performed \citep{Cekli2010}.

The field of pulsating pipe flow received significantly more attention, presumably because of its clear industrial and biophysical relevance, see e.g.\ \citet{Womersley1955,Shemer1985, Mao1986,Lodahl1998,He2009}, and many others. In most studies, like in the present study, an oscillatory flow was superimposed on a steady current.
Depending on the relative strength, the system was either `current-dominated' or, for strong oscillations, `wave-dominated', the  majority of the studies being current-dominated \citep{Manna2012}. 
\red{For many cases it was found that pulsations increase the critical Reynolds number \citep{Sarpkaja1966,Yellin1966}, and, an initially turbulent flow {\it can} relaminarize when a periodic forcing is applied \citep{Ramaprian1980,Shemer1985}.} In most studies the Reynolds number of the imposed oscillatory flow however was close to the laminar-turbulent transition \citep{Lodahl1998}, thus, even if the steady current was fully turbulent, the oscillation was not.

Periodically driven turbulence also includes studies in a number of different well-known and canonical closed-flow geometries, such as Rayleigh-B\'enard convection \citep{jin2008,Sterl2016}, and von K\'arm\'an flow \citep{cad03}. In these systems the forcing was periodically varied over time, with the variations being of $O$(10\%) of either the average forcing or the energy input.

The main observations made in the studies on {\it sinusoidal driven} turbulence were similar regarding the global response of the system \citep{hey03a,hey03b,cad03,kuc06,chi13}. The periodic driving is governed by the Womersley number $W\!o = L\sqrt{ \Omega/\nu}$, which can be seen as the square root of the dimensionless modulation frequency. Here, $L$ is a characteristic length-scale, $\nu$ the kinematic viscosity, and $\Omega $ the angular oscillation frequency. In the limit of extremely small Womersley numbers, the flow can fully respond to the changes, meaning that the flow behaves quasi-stationary. In this regime, no phase delay $\Phi_{delay}$ between the response and the modulation is observed, and the response amplitude is identical to the modulation amplitude. As the Womersley number is increased, the fluid system cannot follow the changing BC: the response amplitude decreases and a phase delay between input and response is observed. In the extreme case of infinite Womersley numbers, the response amplitude vanishes and a phase delay can no longer be defined.

In this manuscript, we study the physics of periodically driven turbulence in a Taylor-Couette (TC) apparatus, employing a {\it sinusoidally} driven inner cylinder.  TC flow, i.e.\ the flow of a fluid confined in the gap between two concentric cylinders, is one of the canonical systems  in which the physics of fluids is studied, see e.g.\ the recent reviews by \citet{far14} and \citet{gro16}. It has the advantage of being a closed system with an exact global energy balance \citep*{eck07b}, and due to its simple geometry TC systems can be accessed experimentally with high precision. 

An important difference between pipe flow and TC flow is the way the system is driven. Pulsating pipe flow is driven by a time-dependent pressure difference applied to the system, but the walls remain fixed. Therefore, momentum is transported from the bulk flow to the boundary layers. In TC flow, the (periodic) driving is by the rotation of the cylinders, so that the momentum is transported from the boundary layer to the bulk flow. By periodically driving the inner cylinder we directly modulate the boundary layer, which transports the modulations to the bulk flow, whereas in pipe flow the bulk flow is directly modulated by the applied pressure gradient. Therefore, studying periodically driven Taylor-Couette turbulence sheds light on the role of the boundary layers in transporting these modulations. \red{Further important differences are the presence of curvature effects and centrifugal forcing in TC, which are clearly absent in pipe flow.}
Apart from several recent studies which focussed on the {\it decay} of turbulent TC flow \citep{ost14jfmr,Verschoof2016,ost17}, or time-dependent driving close to the low Reynolds number Taylor-vortex regime \citep{ahl87,wal88,bar89,gan94,bor10},  to our knowledge no work has been conducted so far on TC turbulence with time-dependent driving.

The outline of this article is as follows. We start by explaining the experimental method in \S 2. The results, in which we present the response of the flow, are shown in \S 3. Finally, we conclude this paper in \S 5. 

\begin{figure}
\centering
\begin{subfigure}{.4\textwidth}
  \centering
  \includegraphics[width=1\linewidth]{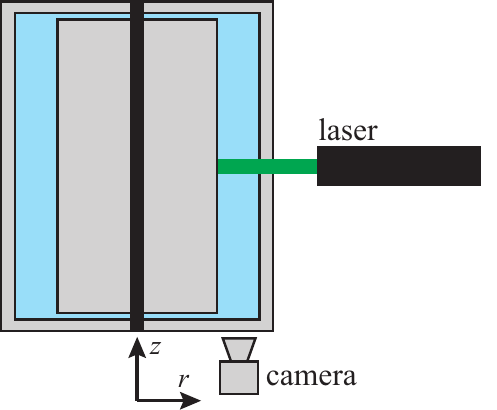}

\end{subfigure}%
\begin{subfigure}{.6\textwidth}
  \centering
  \includegraphics[width=1\linewidth]{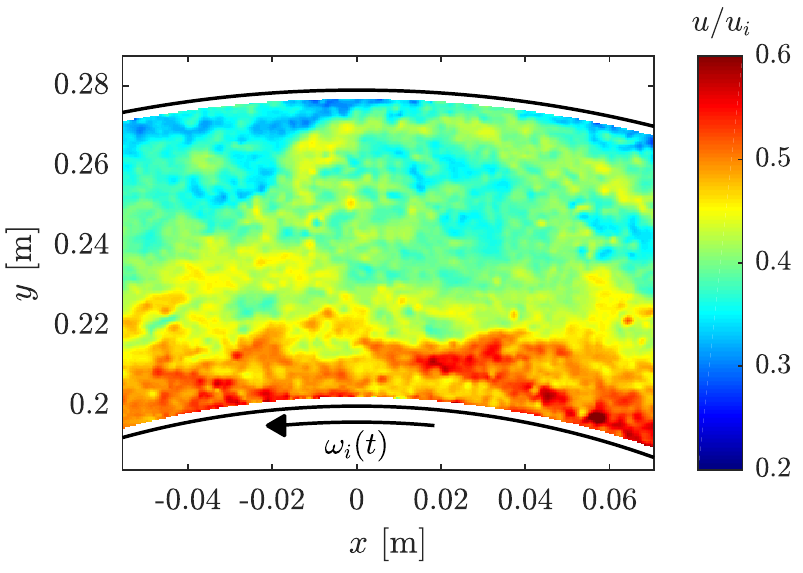}

\end{subfigure}

\caption{Schematic of the vertical cross-section of the T$^3$C facility. The laser illuminates a horizontal plane ($r,\theta$) at midheight ($z = l/2$)  for all PIV measurements. The flow is imaged from the bottom with a high resolution sCMOS camera to obtain the velocity components $u_{\theta}$ and $u_r$ in the ($r,\theta$) plane. On the right we show a typical instantaneous flow field, as measured with PIV. Here we show $u = \sqrt{u_r^2 + u_{\theta}^2}$ \red{normalized with the inner cylinder velocity $u_i$}, for the case with $W\!o = 44.3$, $\Phi = 2.17$ radians and an instantaneous Reynolds number of $\Rey_i = 5.4 \times 10^5.$}
\label{fig:fig1}
\end{figure}

\section{Method}
In this study, we restrict ourselves to the case of inner cylinder rotation, while keeping the outer cylinder at rest. The inner cylinder rotation is set to
\begin{equation}
f_i(t) = \langle f_i \rangle_t \left(1  + e \sin(2 \pi t/ T) \right),
\end{equation}
in which $f_i(t)$ is the rotation rate of the inner cylinder at time $t$ and $T=2\pi/\Omega$ is the period of the modulation. The time $t$ is related to the phase $\Phi$ by $\Phi = 2 \pi t/T$.  We here chose to study the current-dominated regime. To do so, the modulation amplitude is set to $e=0.10$ throughout this work, so that the mean flow is one order of magnitude larger than the induced modulation.  The time-averaged rotation rate $\langle f_i \rangle_t$ is set to $\langle f_i \rangle_t=5$~Hz, resulting in a time-averaged Reynolds number of $\langle \Rey_i \rangle_t = \langle u_i \rangle_t d / \nu = 2 \pi \langle f_i \rangle_t r_i d / \nu = 5 \times 10^5$. In this equation, $u_i=2\pi f_i r_i$ equals the velocity of the inner cylinder with radius $r_i$, $\nu$ is the kinematic viscosity and $d$ is the gap width between the cylinders. Here, we are in the so-called `ultimate turbulence' regime, in which both the bulk flow and boundary layers are fully turbulent \citep{kra62,cha97,gro11,hui12}. The strength of the modulation, which can be estimated as $\Delta\Rey_i \equiv e\langle \Rey_i \rangle_t = 5\times 10^4$, is such that the system is well in the ultimate regime at all times. We varied the modulation period $T$ from 180~s down to 3~s.  The modulation period can be made dimensionless, resulting in the Womersley number, which is defined as
\begin{equation} W\!o = d \sqrt{2 \pi /(T \nu)}.
\end{equation}
See table \ref{table} for all experimental parameters. The Womersley number is connected with the Stokes boundary layer thickness $\delta = 2\pi \sqrt{2 \nu T / (2 \pi)}$, which, in its dimensionless form $\tilde{\delta}=\delta / d =\sqrt{8}\pi/W\!o $, is proportional to the inverse of the Womersley number. The modulation frequency was limited by the power of the motor needed to accelerate and decelerate the mass of the inner cylinder (160 kg). Due to vibrations in the system, higher order statistics cannot be measured. We then simultaneously measured the rotational speed of the inner cylinder $f_i(t)$ and the fluid velocity  by using non-intrusive Particle Image Velocimetry (PIV).

\begin{table}
\begin{center}
\begin{tabular}{ccccc}
$\langle \Rey_i \rangle_t$ & $\Delta\Rey_i$& $T$ [s] & $W\!o$ & $\tilde{\delta}$ \\
\hline
$5\times 10^5$ & $5\times 10^4$ &3 & 114.3 & 0.078\\
$5\times 10^5$ & $5\times 10^4$ &5 & 88.6 & 0.100\\
$5\times 10^5$ & $5\times 10^4$ &10 & 62.6& 0.142\\
$5\times 10^5$ & $5\times 10^4$ &20 & 44.3 & 0.201\\
$5\times 10^5$ &$5\times 10^4$ & 30 & 36.2 & 0.246\\
$5\times 10^5$ &$5\times 10^4$ & 60 & 26.6 & 0.348\\
$5\times 10^5$ &$5\times 10^4$ & 90 & 20.9 & 0.426\\
$5\times 10^5$ &$5\times 10^4$ & 180 & 14.8 &0.602
\end{tabular}
\end{center}
\caption{Experimental details of the measurements. In all measurements the time-averaged Reynolds number as well as the modulation strength as kept identical. By changing the modulation period $T$, we consequently change the Womersley number $W\!o$. In the last column, we show the normalized Stokes boundary layer thickness $\tilde{\delta} = \delta / d$.}
\label{table}
\end{table}

The experiments were performed in the Twente Turbulent Taylor-Couette (T$^3$C) facility \citep{gil11a}, as shown schematically in figure \ref{fig:fig1}. The apparatus has an inner cylinder with a radius of $r_i=200$ mm and a transparent outer cylinder with a radius of $r_o=279.4$ mm, resulting in a radius ratio of $\eta=r_i/r_o=0.716$, a gap width $d=r_o-r_i=79.4$ mm. The height of the setup is $l=927$ mm, giving an aspect ratio of $\Gamma = l/d=11.7$. As working fluid we use water with a temperature of $T= 20~^{\circ}$C, which is kept constant within 0.2 K by active cooling through the end-plates of the setup. More experimental details of this facility can be found in \citet{gil11a}.

The PIV measurements were performed in the $r-\theta$ plane at mid-height ($z=l/2$) using a high-resolution camera operating at 15~fps (pco.edge camera, double frame sCMOS, 2560$\times$2160 pixel resolution). We illuminate the flow from the side with a horizontal laser sheet, as shown in figure \ref{fig:fig1}. The used laser is a pulsed dual-cavity 532 nm Quantel Evergreen 145 Nd:YAG laser. We seeded the water with 1-20 $\mu$m fluorescent polyamide particles. We calculate the Stokes number which equals $Stk = \tau_p / \tau_{\eta}= 0.0019 \ll 1$. Furthermore, the mean particle radius is roughly 6 times smaller than our Kolmogorov length scale, thus we can be sure that the particles faithfully follow the flow. The images are processed with interrogation windows of 32 $\times$ 32 pixel with 50\% overlap, resulting in $u_\theta(r,\theta,t)$ and $u_r(r,\theta,t)$. 
We were unable to measure close to the cylinders due to the strong laser light reflections.

To compare our experiments in highly turbulent flow with the laminar case, we numerically solved the response of the flow. We therefore solved the partial differential equation
	\begin{equation}
	\frac{\partial u_{\theta}}{\partial t} = \nu \left [\frac1r \left ( \frac{\partial}{\partial r} \left (r \frac{\partial u_{\theta}}{\partial r}\right ) \right ) - \frac{u_{\theta}}{r^2} \right ], 
	\end{equation}
	which is the time-dependent Navier-Stokes equation in cylindrical coordinates for the azimuthal direction under the assumptions of i) no azimuthal and axial derivatives, ii) $u_r=0$ and $u_z=0$, so that $\vec{u}(r,\theta,z,t) = u_{\theta}(r,t) \hat{e}_{\theta}$.
	As initial condition we used the steady-state laminar flow profile, i.e. $u_{\theta}(r,t=0) = \frac{1}{1-\eta^2} \left(\frac{r_i^2 \omega_i}{r} - \omega_i \eta^2 r \right)  $. As time-dependent boundary conditions we set $u(r_i,t)=\omega_i r_i \left(1+0.1 \sin(2 \pi t/T) \right)$, and the outer cylinder is stationary, i.e.\ $u(r_o,t)=0$. We run the computation for 40 periods, so that all transient effects are gone.

\begin{figure}
\centering
\includegraphics{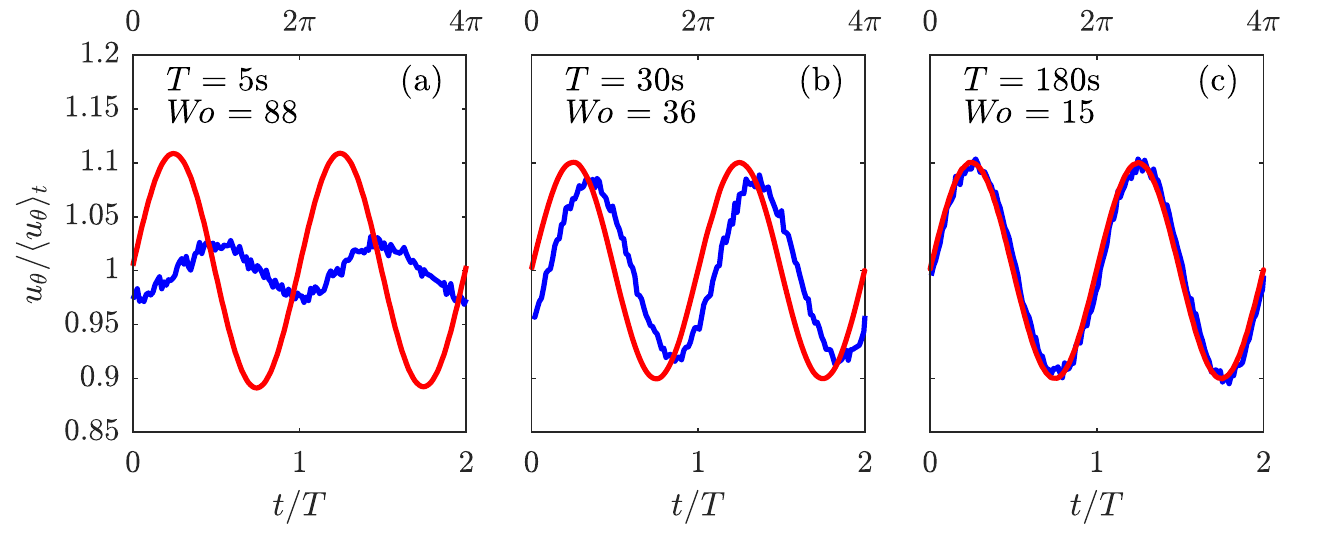}
\caption{  \protect \redline~Normalized azimuthal velocity of the sinusoidally driven inner cylinder $u_i / \langle u_i \rangle_t$. \protect\blueline~Normalized azimuthal velocity $u_{\theta}/ \langle u_{\theta} \rangle_t$  at mid-gap. Three Womersley numbers are shown, namely (a) $W\!o=88$, (b) $W\!o=36$, and (c) $W\!o=15$. The velocity is radially averaged between $0.3\leq \tilde{r}  \leq 0.7$. On the top x-axis, we show the phase $\Phi$ of the modulations in radians.} 
\label{fig:fig2}
\end{figure}
\begin{figure}
\centering
\includegraphics[scale=1]{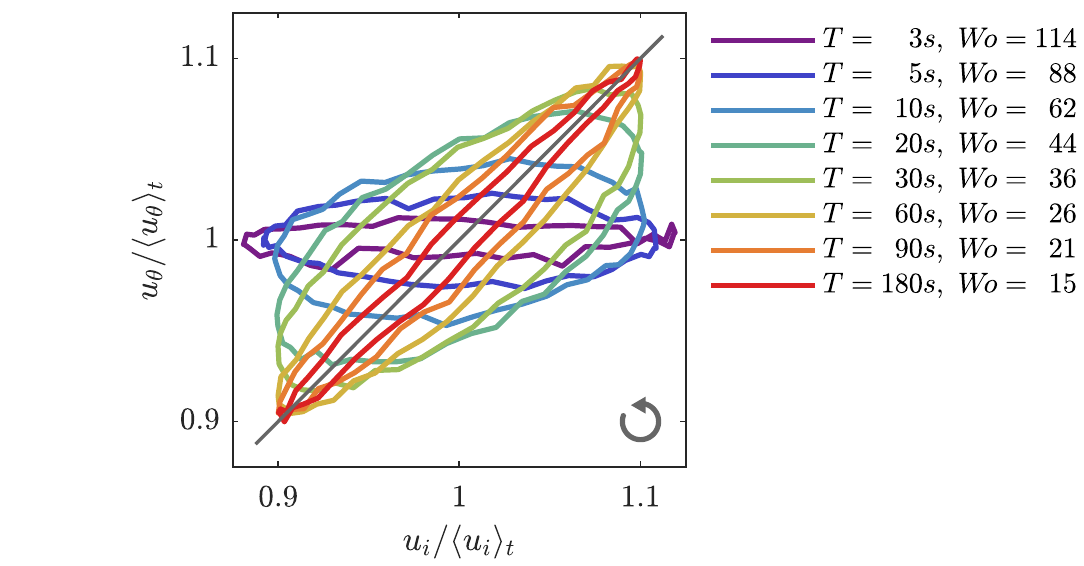}
\caption{Phase averaged normalized azimuthal mid-gap flow velocity $u_{\theta}/ \langle u_{\theta} \rangle_t$ as a function of normalized driving velocity of the inner cylinder $u_i / \langle u_i \rangle_t$. We show the result for all measured Womersley numbers $W\!o$. The velocity is radially averaged between $0.3\leq \tilde{r} \leq 0.7$. The solid grey line corresponds to the quasi-stationary case $u_{\theta}/ \langle u_{\theta} \rangle_t =u_i / \langle u_i \rangle_t$. The arrow at the bottom right indicates the direction of the cycles.}
\label{fig:fig3}
\end{figure}

\section{Results and analysis}
\subsection{Velocity response}
In figure \ref{fig:fig2} we show the normalized driving and response of the mid-gap flow velocity $u_\theta(\tilde{r}=0.5,t)$ for three different modulation periods. The radius is non-dimensionalized as $\tilde{r} = (r-r_i)/d$, so that $\tilde{r}=0$ corresponds to the inner cylinder and $\tilde{r}=1$  to the outer one. We non-dimensionalize both velocities by their time-averaged value, so both lines meander around $1$. For all oscillation periods, the mid-gap flow velocity oscillates with the same period $T$ as the driving. The amplitude and phase delay of the response depend on the driving period. For the larger modulation periods $T$, $u_{\theta}$ responds nearly in phase with the same amplitude as the driving. For smaller modulation periods, the response amplitude decreases and a phase delay is observed, just as in prior studies \citep{hey03a,hey03b,cad03,Kuczaj2008,Hamlington2009}.

\begin{figure}
\centering
\includegraphics{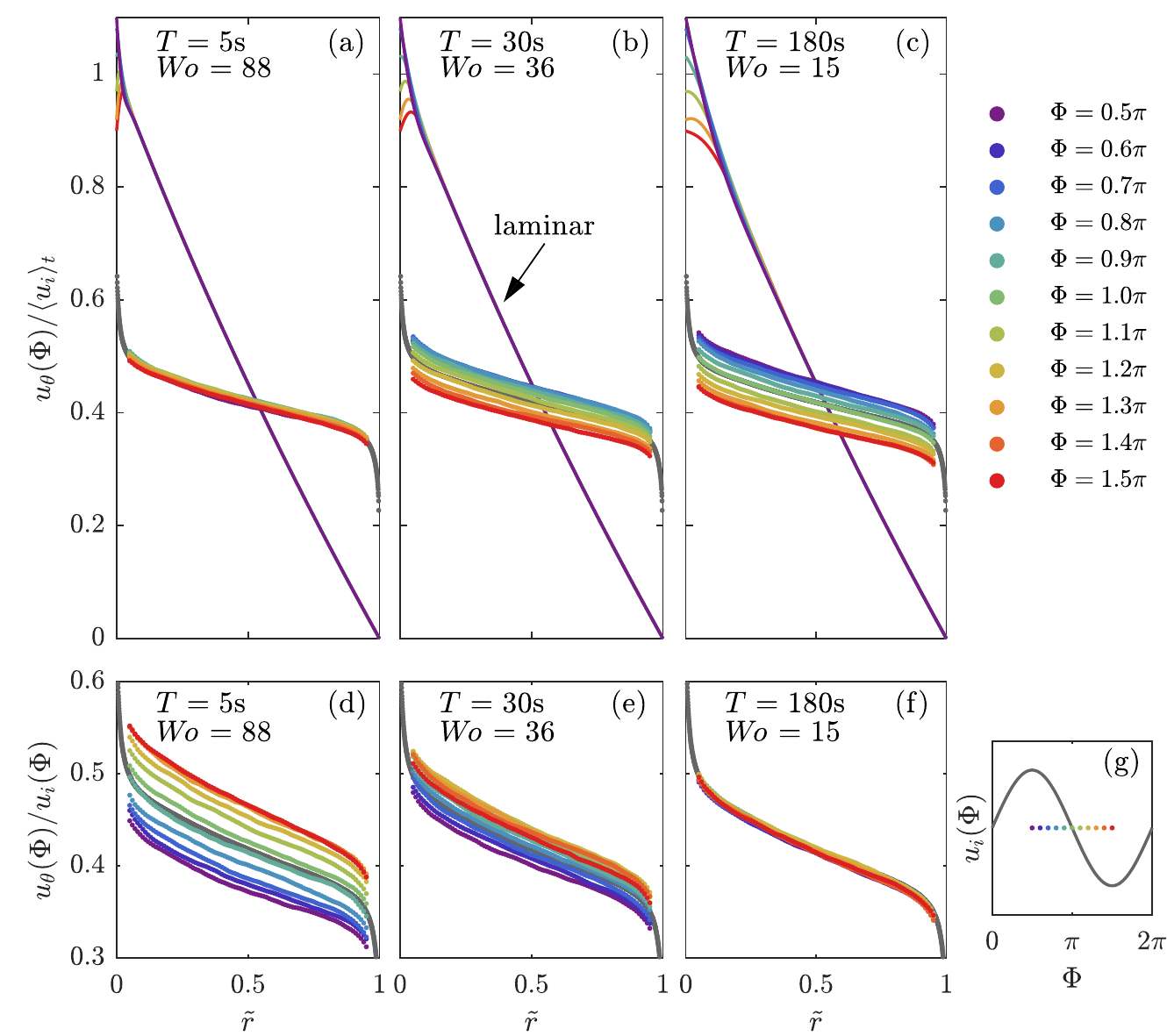}
\caption{Azimuthal velocity profiles as a function of dimensionless radius $\tilde{r}$. All data is phase-averaged and normalized. 
\protect\\ {\bf Top row (a-c)} $u_{\theta}(\Phi)$ is normalized with the time-averaged inner cylinder velocity $\langle u_i \rangle_t =6.3$~m/s, i.e.\ the same constant value for all lines. A collapse of all lines indicates that the response amplitude is small, as is observed for large $W\!o$, see figure (a). Furthermore, we show the response of laminar flow to the modulation, calculated numerically (see method section). \protect\\ 
{\bf Bottom row (d-f)} $u_{\theta}(\Phi)$ is normalized by the instantaneous inner cylinder velocity at phase $\Phi$, i.e.\  $u_{i}(\Phi)$ (a value between $u_i(0.5 \pi) = 6.9$~m/s and $u_i(1.5 \pi) = 5.7$~m/s). A collapse of all lines indicates that the system behaves quasi-stationary, as can be seen for small $W\!o$ in figure (f).\protect\\
The solid \red{grey} lines show the azimuthal velocity profile for $\Rey_i = 5\times 10^5$ for the non-modulated, stationary case (data from \cite{hui13}). \protect\\
{\bf Bottom right (g)} The azimuthal velocity $u_{\theta}(\Phi)$ is shown for a series of phases of the modulation; here we show data for phases between $0.5\pi \leq \Phi \leq 1.5\pi$, i.e.\ half of a modulation cycle, as shown in this inset. See also figure \ref{fig:fig2} for the definition of phase $\Phi$. }
\label{fig:fig4}
\end{figure}

A different representation of a modulation cycle is depicted in figure \ref{fig:fig3}. Here we plot the data from figure \ref{fig:fig2} parametrically as a function of $\Phi$. A fully quasi-stationary cycle completely follows the grey line, in which $u_{\theta}/ \langle u_{\theta} \rangle_t =u_i / \langle u_i \rangle_t$. The $W\!o=15$ measurement is close to this line. The deviation from this line, which indicates a phase delay, increases for smaller modulation periods. 

To study whether the flow responds similarly over the gap width, we extend the analysis from figure \ref{fig:fig2} to the entire radius, see figure \ref{fig:fig4}. In the top row, the data is normalized by $\langle u_i \rangle_t =  2 \pi \langle f_i \rangle_t r_i = 6.3$~m/s, i.e. the same constant for all measurements. The better all lines collapse, the smaller the response amplitude is. For the bottom row, we chose to normalize with $u_i(\Phi)=2 \pi r_i \langle f_i\rangle_t[1+e\sin(\Phi)]$, i.e. the inner cylinder velocity at the corresponding phase in the modulation. Here, when all lines collapse, the modulation is slow enough for the flow to react to the modulation, i.e. the system is in a quasi-stationary state. For comparison,  the azimuthal velocity profile for the non-modulated case is shown as a grey line \citep{hui13}. Figure \ref{fig:fig4}(a) and (f) depict the most extreme cases. Furthermore, we show the laminar flow response in the top row. In figure \ref{fig:fig4}(a), the azimuthal velocity of the flow is almost constant over a modulation cycle, and therefore $u_{\theta}(r,\Phi)$ is close to the non-modulated statistically stationary solution for $f_i=5$~Hz; the flow cannot adapt to the quick changes of the inner cylinder. For larger Womersley numbers, the opposite is the case, see figure \ref{fig:fig4}(f). Here, for every phase $\Phi$, the azimuthal velocity profile is identical to the statistically stationary solution for $f_i(\Phi)$. This behaviour is surprisingly constant over the entire radius. We note that it might appear as if the correct boundary conditions are not met. However, as shown in \citet{hui13}, the boundary layer at the studied Reynolds number is too thin to resolve from the current measurements.

The laminar flow response is completely different as compared to the measured turbulent case. First, the response in the flow is restricted to a thin layer close to the inner cylinder wall. Calculating the thickness of the Stokes boundary-layer, although slightly off due to the presence of the outer cylinder and a cylindrical coordinate system, gives a similar result, i.e.\ $\tilde{\delta}(W\!o=88)= 0.10$, $\tilde{\delta}(W\!o=36)= 0.24$, and $\tilde{\delta}(W\!o=15)= 0.60 $ (see table \ref{table}).  Second, the response is radius-dependent, as is also known from Stokes oscillating plate theory, as the response decays exponentially with increasing distance from the oscillating wall. These observations highlight how turbulent mixing enhances the transport of the modulation over the entire radius.

\begin{figure}
\centering
\includegraphics[scale=1]{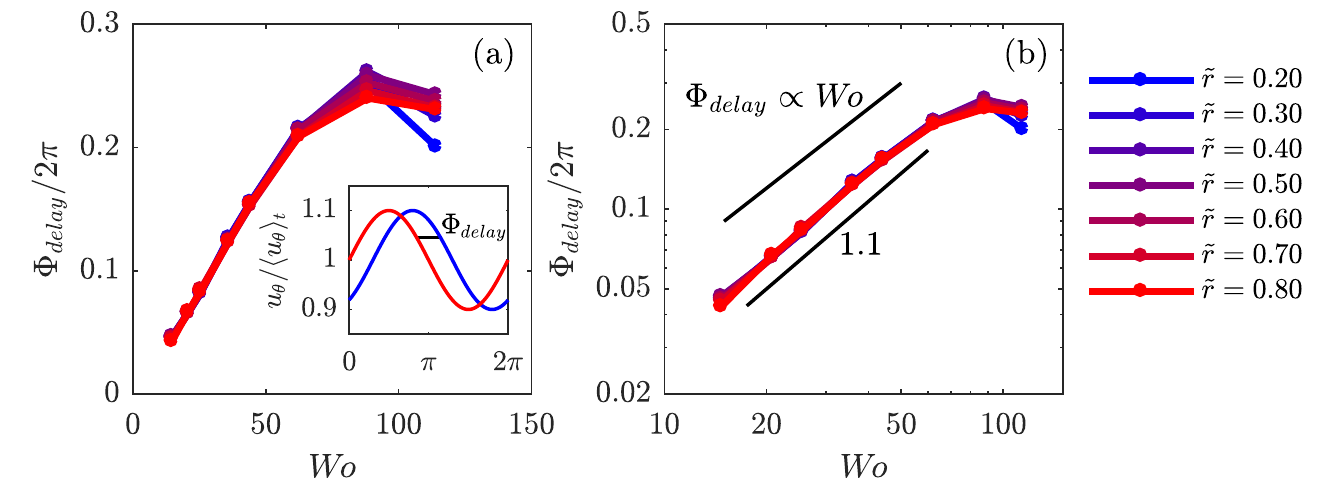}
\caption{The delay between the driving modulation and the fluid velocity response as a function of Womersley number $W\!o$. The delay $\Phi_{delay}$  is normalized with $2\pi$ of the modulation. The phase delay is calculated for a number of radii, not showing much difference.  The figures show the same data in linear scale (a) and logarithmic scale (b). The results are radially binned within $\tilde{r} \pm 0.025$. The inset in figure a) shows how the phase delay $\Phi_{delay}$ is defined. $\Phi_{delay}$ is calculated by cross-correlating both signals. We included the scaling of the response for laminar flow, which equals $\Phi_{delay} \propto W\!o$.}
\label{fig:fig5}
\end{figure}

\subsection{Phase delay}
Up to now the conclusions drawn from figures \ref{fig:fig2}, \ref{fig:fig3}, and \ref{fig:fig4} were only qualitative. Here, we quantify the phase shift and amplitude response for the turbulent case. We extract the phase delay $\Phi_{delay}$ between the modulation and the response by cross-correlating $u_i(t)$ and $u_{\theta}(t)$. We detect the first peak in $u_i \star u_{\theta}(\tau)$, and obtain the phase delay by fitting a Gaussian function through this peak and its two neighbouring points, to obtain the peak with increased accuracy. As visible in figure \ref{fig:fig5}, at large modulation periods, the phase delay is small, as we already qualitatively concluded from figure \ref{fig:fig2}. As the Womersley number increases, the bulk flow cannot follow the changing BCs anymore and it responds with an increasing delay. Within this approximation, \citet{hey03a} calculated, and \citet{cad03} experimentally found, that the phase delay has a linear dependence on the modulation frequency, i.e.\  $\Phi_{delay} \propto W\!o^2$.  
We do not observe a similar behaviour, however. The results in the aforementioned studies, which both study homogeneous and isotropic turbulence (HIT), are significantly different than what we observe in our Taylor-Couette setup, which cannot be regarded as HIT \citep{hui13pre}.

As visible in figure \ref{fig:fig5}b, in this work the dependence of $\Phi_{delay}$ is better described by an effective power law over a range of larger values of $W\!o$, with $\Phi_{delay} \propto W\!o^{1.1}$.  For the laminar case, the phase lag in the Stokes boundary layer problem is calculated as $\Phi_{delay} =   \sqrt{2} \tilde{r} W\!o $. The exponent $1.1$ is close to the value of the laminar flow response, in which there is a linear dependance between the Womersley number and the phase delay.   The phase lag saturates at around $\Phi_{delay}=\pi/2$, similar to what is known in pulsating pipe flow \citep{Womersley1955,Shemer1985} and in e.g.\ periodically forced harmonic oscillators.

We now come to the spatial dependence of the response. Intuitively, one expects an increasing phase delay further away from the modulated wall. Surprisingly, this is not the case. Apparently, the turbulent mixing of this highly turbulent flow prevents the system from having a range of phase delays  over the radius, given the fact that the modulation has been ``passed on'' from the boundary layer to the bulk flow. This can be explained by calculating a characteristic timescale $\tau_{bulk}$ for the movement from  the inner to the outer cylinder, using the Reynolds wind number $Re_w = \sigma(u_r) d / \nu$, in which $\sigma(u_r)$ is the standard deviation of the radial velocity. We estimate $\tau_{bulk} = d/\sigma(u_r) = d^2 / Re_w \nu$. $Re_w$ for the corresponding $\langle Re_i \rangle_t=5\times 10^5$ is known from \cite{hui12}, resulting in a $\tau_{bulk}=0.27$ s. As long as $\tau_{bulk} \ll T$, the radial dependence of the phase delay and amplitude should be negligible, in agreement with our observations. Such small periods $T$ are unfortunately not accessible experimentally due to the moment of inertia of the cylinders.

\begin{figure}
\centering
\includegraphics[scale=1]{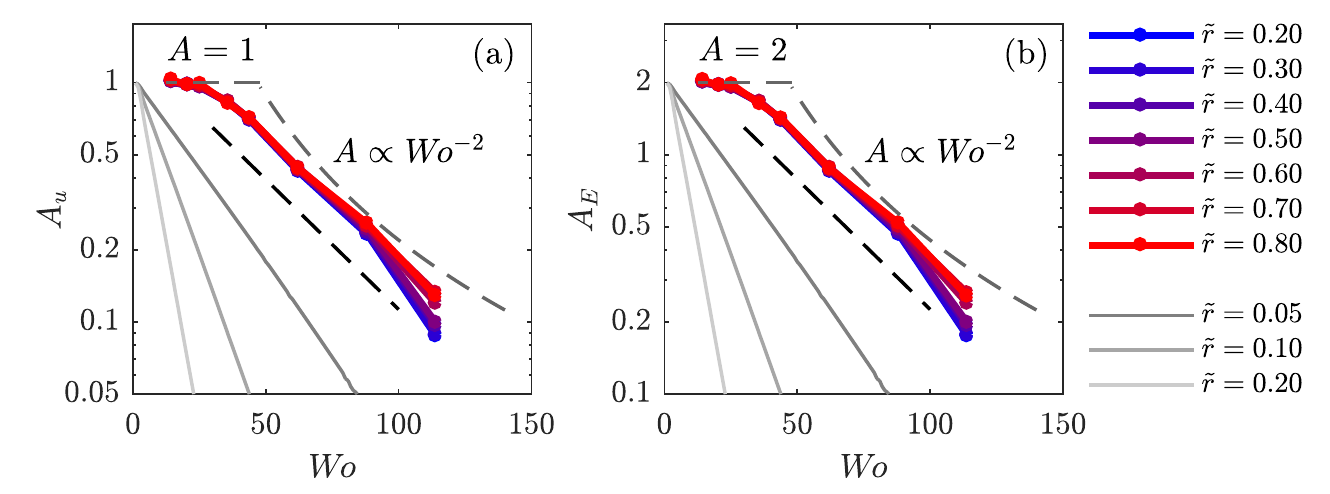}
\caption{\red{Amplitude response as a function of the Womersley number $W\!o$ for various dimensionless radii. The coloured lines represent our measurements, and the solid grey lines are numerically calculated laminar flow responses. (a) The response amplitude of the velocity $A_u$ and (b) the response amplitude of the energy  $A_E$. The experimental results are radially binned between $\tilde{r} \pm 0.025$. The dashed grey lines show the scalings of $A$ as predicted by \citet{hey03a}. We included the laminar responses, shown in solid grey lines. A number of radii are included, to highlight  the 
dependence  on the radius, which does not exist in the well-mixed  turbulent case. 
The effective slope of the measurements $A \propto e^{-0.025 W\!o}  $
is shown in dashed black. This would correspond to  the slope of the laminar flow response at  $\tilde{r}\approx0.035$.}}
\label{fig:fig6}
\end{figure}

\subsection{Amplitude response}
We calculate the amplitude $A$ of the response for both the velocity and kinetic energy, which is defined as $E = \frac12 \vec{u} \cdot \vec{u} \approx \frac12 u_{\theta}^2$. Following the approach of \cite{hey03a}, the local oscillating response of the velocity and energy is calculated as
\begin{equation}
\begin{aligned}
\Delta_u(t) &= \frac{u_{\theta}(t)}{\langle u_{\theta} \rangle_t} - 1, ~~\text{and} \\
\Delta_E(t) &= \frac{E(t)}{\langle E \rangle_t} - 1. \label{eq:eq1}
\end{aligned}
\end{equation}

We average $\Delta_{u}(t)$ and $\Delta_{E}(t)$ radially and azimuthally, and make the ansatz that $ \Delta_{u,E}(t) $ can be described as:
\begin{equation}
\Delta_{fit}(t) = eA(T)\sin(2\pi t/T + \Phi_{delay}).  \label{eq:eq2}
\end{equation}
$\Delta_{fit}(t)$ is fitted to $\Delta(t)$ with $A(T)$ as sole fitting parameter. $\Phi_{delay}$ is not a fitting parameter, as it is calculated using cross-correlation, see figure \ref{fig:fig5}. In the case of slow, quasi-stationairy modulations, the amplitude response of the azimuthal velocity can be calculated from equations (\ref{eq:eq1}), namely  $A_u=\left(\frac{(1+e)}{1}-1\right)/e = 1$. Strictly speaking is it impossible to describe the kinetic energy with a sinusoidal function, as it has a squared dependence on the velocity, but, as $e$ is small a sine wave can be used within the assumption of a linear response. However, the calculation of $A_E$ in the quasi-stationary case is less straight-forward, as the response amplitude varies over the sine wave. We calculate $A_E^{max} = \left((1+e)^2-1\right)/e = 2.1$ and $A_E^{min} = \left((1-e)^2-1\right)/{\text{-}e} = 1.9$  as the two extremes, leading to a phase-averaged value of $A_E=2.0$. Both response amplitudes will vanish in the limit of infinitely fast modulations, i.e.\ $W\!o \to \infty$ implies that $A_{u,E} \to 0$.

As figure \ref{fig:fig6} clearly shows, the fluid completely follows the imposed modulation at larger modulation periods, i.e.\  amplitude responses of $A_u=1$ and $A_E=2$ are observed, which corresponds to our expectations. For smaller modulation periods, the response amplitude decreases. We do not observe clean power laws, as predicted assuming HIT by \citet{hey03a} and \citet{cad03} shown as dashed lines. The response of the flow can better be described by an exponential function, as indicated by the solid black line. This is in line with the laminar flow response, in which the amplitude of the response also is an exponential function of the Womersley number and the distance to the modulated wall. \red{Note that, in contrast to the turbulent case, the amplitude response of the laminar case depends on the radius.}

Similar to the phase delay between modulation and response, also in the response amplitude we do not observe any trend over the radius. Here, one could expect a decreasing $A$ at higher radii, i.e.\ further away modulated wall. Because of the no-slip condition at the wall, the values of $A$ and $\Phi_{delay}$ directly at the wall are fixed, i.e. $A_u(r_i) = 1$ and $\Phi_{delay}(r_i)=0$. At the outer cylinder, $A_u(r_o) = 0$, hence $\Phi_{delay}(r_o)$ cannot be defined. Clearly, the boundary layers play a pivotal role in transferring perturbations and modulations to the bulk of the flow.

\section{Summary and conclusions}
To conclude, we studied periodically driven Taylor-Couette turbulence. We drove the inner cylinder sinusoidally, and measured the local velocity using PIV. Consistent with earlier studies and theoretical expectations, we observe a phase delay and declining velocity response as we increase the Womersley number. Most surprisingly, we did not observe a radial dependence of the phase delay in the bulk of the flow, nor of the amplitude response, in contrast to the expectation one might have that there could be a larger influence of the modulation on the flow close to the modulated wall. Apparently, a radial dependence of $A$ and $\Phi_{delay}$ is prevented by the strong mixing in this turbulent flow. Even though we did not measure directly in the boundary layers, their vital importance in transferring modulations to the bulk flow is evident. This contrasts our numerical results for laminar flow, where a strong radial dependence is observed, and the response of the flow is confined to a thin layer close to the modulated wall. Therefore it is even more remarkable that the scaling relations of both the phase delay and the amplitude response are similar to what had been found for laminar flows. 

To further study this interesting phenomenon, direct numerical simulations are necessary to cover the extremely high Womersley number range, which is inaccessible in experiments. Using such data, it will be possible to study the interplay between the modulated cylinder, the boundary layers and the bulk in more detail, as the entire flowfield will then be available. Another domain of ``terra incognita'' is the study of modulations with larger amplitude. Here, we limited ourselves to a modulation amplitude of $e = 0.1$. Larger values induce non-linear effects, and linear response type assumptions such as those made in equations (\ref{eq:eq1}) and (\ref{eq:eq2}) will then not be valid anymore.

\begin{acknowledgments}
We would like to thank Gert-Wim Bruggert and Martin Bos for their continual technical support over the years. We thank Pim Bullee, Dennis Bakhuis, Pieter Berghout, and Rodrigo Ezeta for various stimulating discussions. This work was financially supported by the NWO-TTW, NWO-I, and an ERC grant.
\end{acknowledgments}

\end{document}